\documentclass[letter]{aa} 

\usepackage{graphicx}
\usepackage{txfonts}
\usepackage{ulem}
\usepackage{subfigure}
\usepackage{caption,subcaption}
\usepackage{hyperref}

\begin{document}

        \title{Kolmogorov analysis of JWST deep survey galaxies}
        
        
        \author{N. Galikyan \inst{1,2} 
                \and
                A.A.Kocharyan\inst{3}
                \and
                V.G. Gurzadyan\inst{2,4} \fnmsep\thanks{Email: gurzadyan@yerphi.am}
        }
        
        \institute{National Research Nuclear University MEPhI, Moscow, Russia 
                \and Center for Cosmology and Astrophysics, Alikhanian National Laboratory and Yerevan State University, Yerevan, Armenia 
                \and School of Physics and Astronomy, Monash University, Clayton, Australia 
                \and SIA, Sapienza University of Rome, Rome, Italy
        }
        
        \date{Received XXX; accepted ZZZ}

        \abstract
        {JWST galaxy deep spectral surveys provide a unique opportunity to trace a broad range of evolutionary features of galaxies and the intergalactic medium given the huge distance the photons are propagating. We have analyzed the spectral data of JWST galaxies up to a redshift of around 7 using the Kolmogorov technique, which is an efficient tool for testing the tiny comparative randomness properties of cumulative signals, that is, for distinguishing the contributions of regular and stochastic sub-signals. Our aim is to determine if certain identical spectral features of galaxies have undergone any distortions or systematic evolution across a broad range of redshifts. Our results indicate a change in the spectral properties of the sample galaxies at around $z \simeq 2.7$ at over a 99\% confidence level.  
        }

        
        \maketitle
        
        \section{Introduction}
    
The data produced by JWST on the high-redshift Universe reveal remarkable features regarding early galaxies, star formation, supermassive black holes, and numerous associated aspects (see \citealt{Coo,Sil,Liu,Mar,Run,Sun,Per} and references therein). Certain observational data are seen to challenge the predictions of the standard cosmological model, or at least call for the reconsideration of particular issues. Various data analysis techniques, including those involving artificial intelligence and machine learning \citep{Rob}, are being used to study the ever-increasing amount of high-redshift observational data.

We analyzed JWST spectral data of galaxies up to redshift $z \simeq 7$. We employed the Kolmogorov stochasticity parameter (KSP) approach \citep[][]{K,N,UMN} to analyze the comparative features of the spectral signals of the galaxies. The KSP is an efficient tool for studying the comparative randomness features of cumulative signals and dynamical systems \citep{UMN,MMS,FA,Entr}. Namely, the KSP enables signals to be distinguished based on their composition of regular and random sub-signals, and can thus be applied to physical signals with regular components and foreground stochastic noise. Regarding astrophysical signals, the KSP test has been applied in studies of non-Gaussianities in cosmic microwave background (CMB) data \citep{GK,KSKY}. In particular, the KSP enabled \cite{CS} to draw conclusions regarding the void nature of the Cold Spot in the CMB {\it Planck} sky map; soon after that study, a void aligned with the Cold Spot was detected in an infrared galactic survey \citep{Sz}. The dynamical difference of structures at local and global cosmological scales can lead to different observational signals \citep{GS,GFC1,GFC2}.  The difference in the KSP properties enabled  the CMB signal to be clearly distinguished from the microwave emission of the Galactic disk \citep{GK,KSKY}; this revealed \textit{XMM-Newton} X-ray galaxy clusters \citep{Xray} and gamma sources \citep{gamma}, confirming the efficiency of the approach for a broad class of physical issues. 

We used the KSP test to determine if there are any signal distortions or redshift-dependent variations for JWST galaxies across a broad range of redshifts.

\section{Kolmogorov stochasticity parameter}
\label{sec:ksp}
        
        \cite{N,UMN,MMS,FA} outlined the remarkable efficiency of Kolmogorov's theorem \citep{K} in defining the randomness properties of real-valued sequences. Consider a $\{X_1,X_2,\dots,X_n\}$ sequence of $n$ random variables $X$ ordered in an increasing manner, $X_1\le X_2\le\dots\le X_n$, and its cumulative distribution function (CDF):
        \begin{equation}
        F(x) = P\{X\le x\}. 
        \end{equation}
        One can define an empirical distribution function as
        \begin{eqnarray*}
                F_n(x)=
                \begin{cases}
                        0\ , & x<X_1\ ;\\
                        k/n\ , & X_k\le x<X_{k+1},\ \ k=1,2,\dots,n-1\ ;\\
                        1\ , & X_n\le x\ .
                \end{cases}
        \end{eqnarray*}
        
        \noindent Then, the KSP $\lambda_n$ is defined as  
        \begin{equation}\label{KSP}
                \lambda_n=\sqrt{n}\ \sup_x|F_n(x)-F(x)|\ .
        \end{equation}
        
        \noindent \cite{K} proved that, for any continuous CDF, 
        $$
        \lim_{n\to\infty}P\{\lambda_n\le\lambda\}=\Phi(\lambda)\ ,
        $$
        where $\Phi(0)=0$,
        \begin{equation}
                \Phi(\lambda)=\sum_{k=-\infty}^{+\infty}\ (-1)^k\ e^{-2k^2\lambda^2}\ ,\ \  \lambda>0\ ,\label{Phi}
        \end{equation}
        this limit converges uniformly, and Kolmogorov's distribution, $\Phi,$  is independent of the CDF. It can be shown that, within $0.3\le\lambda_n\le 2.4,$ the KSP acts as an objective degree of randomness \citep{N,UMN}. These features of the Kolmogorov distribution make the KSP test an efficient tool for analyzing the comparative randomness properties of signals.

\section{Dataset} 

The galaxy spectroscopic data of the JWST NIRSpec instrument \citep{JWST,NIRSpec} from the Ultradeep NIRSpec and NIRCam Observations before the Epoch of Reionization (UNCOVER)  survey \citep{UNCOVER1,UNCOVER2} allow spectral signals from the observed galaxies to be inspected across a broad range of redshifts. 

\begin{figure}[h!]
    \centering
    \includegraphics[width=\linewidth]{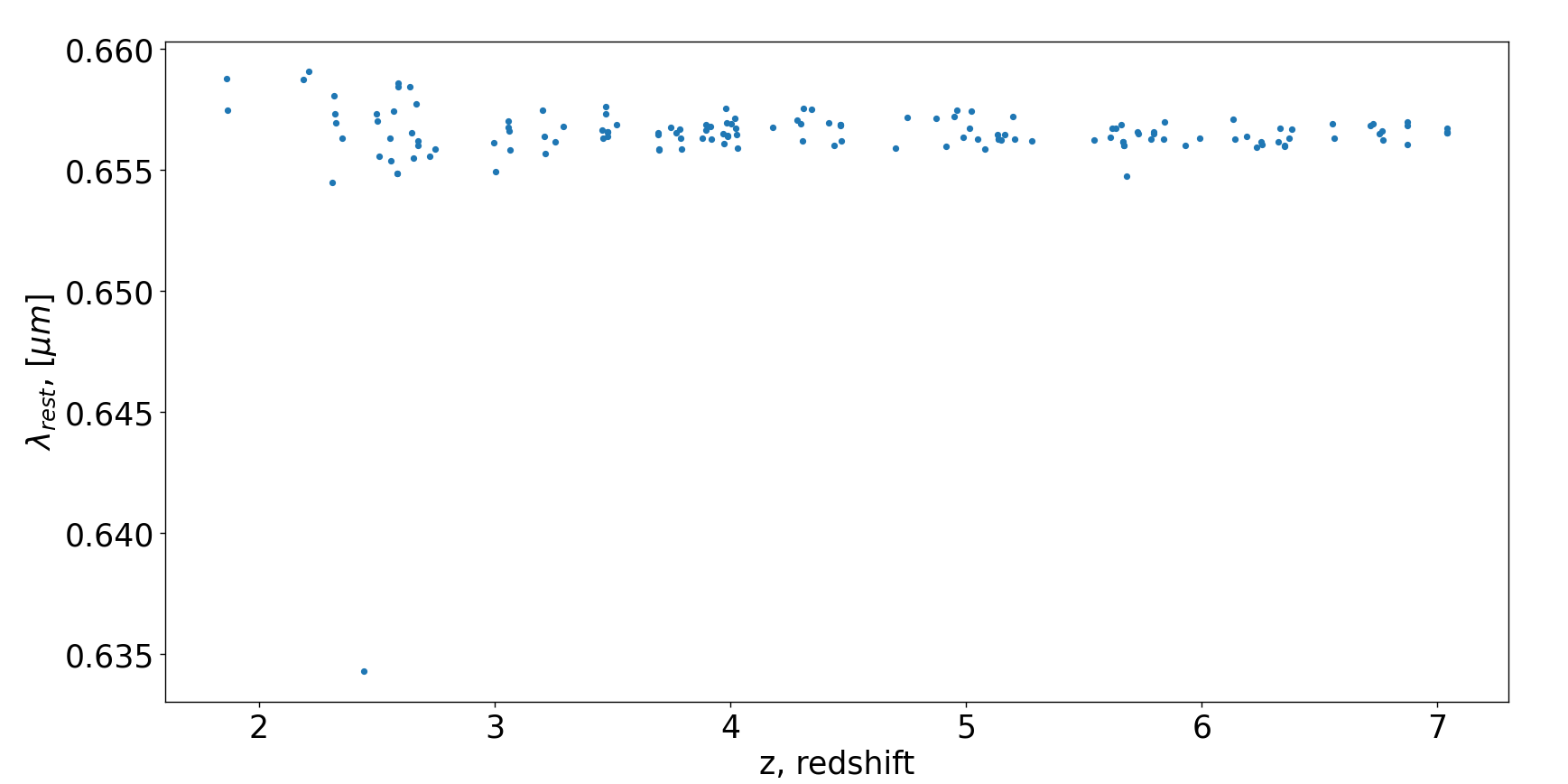}
    \caption{Dependence of the wavelength on the redshift of the galaxies' spectral maxima  in the rest frame.}
    \label{fig:lambda_z}
\end{figure}

\cite{UNCOVER1} divided galaxies  by their redshift fit quality and we performed a KSP analysis of the signals for spectra with redshift quality flags ``secure" and ``solid." Galaxies with a maximum spectrum corresponding to a wavelength nearly equal to $656$~[nm] in the rest frame were selected for the analysis. They cover a broad range of redshifts, $1.86\leq z\leq7.05$, and contain the majority of data points for the sample, including $148$ galaxies. Figure~\ref{fig:lambda_z} shows the dependence of the wavelength corresponding to the maximum spectrum on the redshift of the galaxies.

The procedure described in Sect. \ref{sec:ksp} was performed on the abovementioned wavelength sequence of the galaxies.

\section{Analysis}

We performed a KSP analysis on these galaxy spectral data. To obtain the KSP dependence of the signal at a given redshift, the following procedure was adopted. One thousand nonidentical pairs of uniformly distributed numbers corresponding to redshifts $z_1<z_2$  were generated such that each $z_1<z_2$ interval contained $10$-$20$ galaxies. Then, the KSP was calculated within those interval samples (i.e., $z=\frac{z_1+z_2}{2}$), enabling us to assign  a degree of randomness to a given redshift. We note that although the number of galaxies within the selected redshift intervals is small, the KSP test remains efficient as an indicator of randomness; its capability to detect such characteristics has previously been demonstrated for small data sequences \citep{N}.

\begin{figure}[h!]
\includegraphics[width=\linewidth]{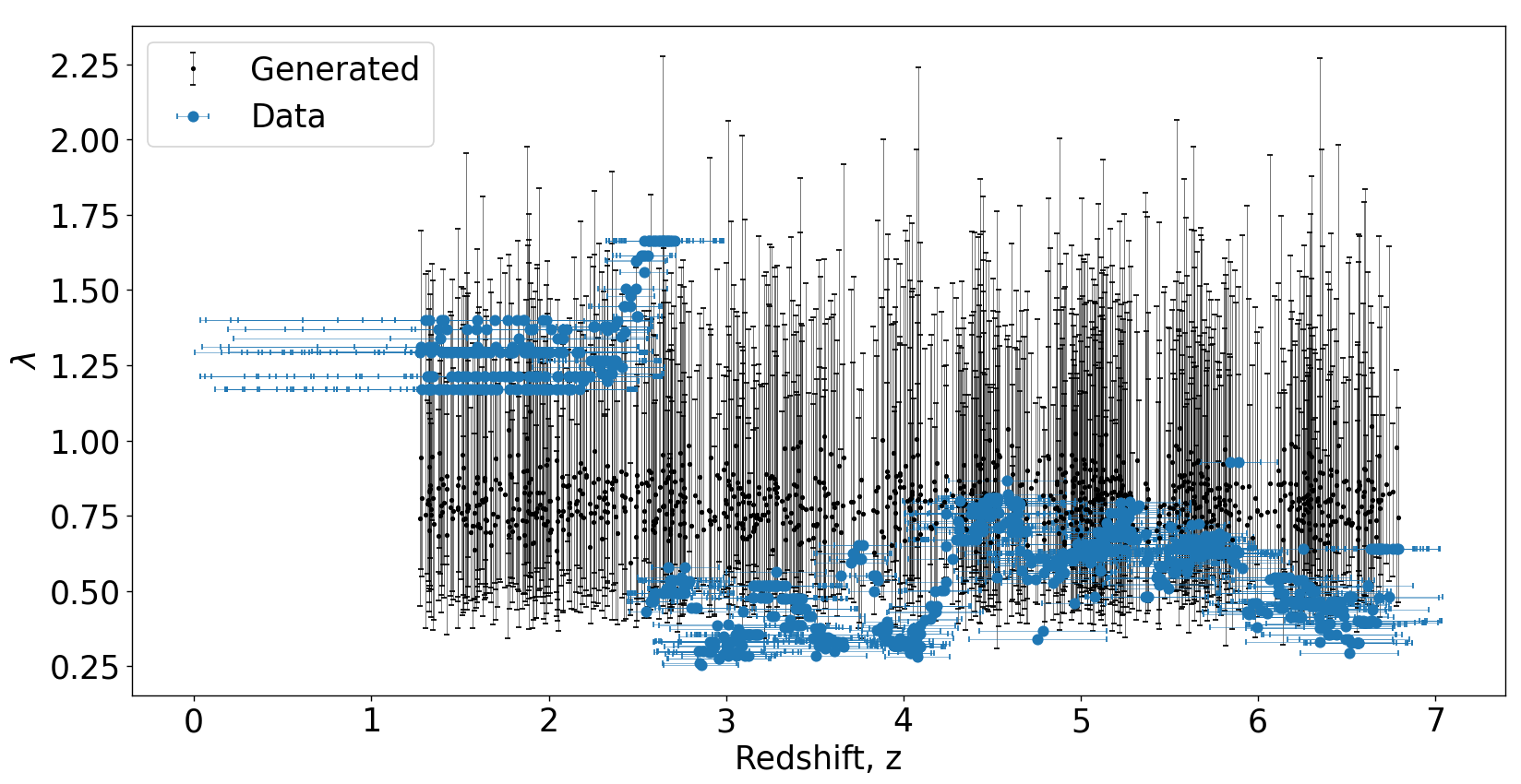}
\caption{Redshift dependence of the KSP $\lambda$ of the galaxy emission corresponding to the maximum of the spectrum. Blue points denote the KSP values calculated for the observed galaxies, and horizontal error bars indicate the selected $[z_1; z_2]$ intervals. Black points indicate the median of the mock data KSP, and the vertical error bars are the 99\% confidence intervals around the medians of the generated data.}
\label{fig:ksp_z}
\end{figure}

\begin{figure}[h!]
    \centering
    \includegraphics[width=\linewidth]{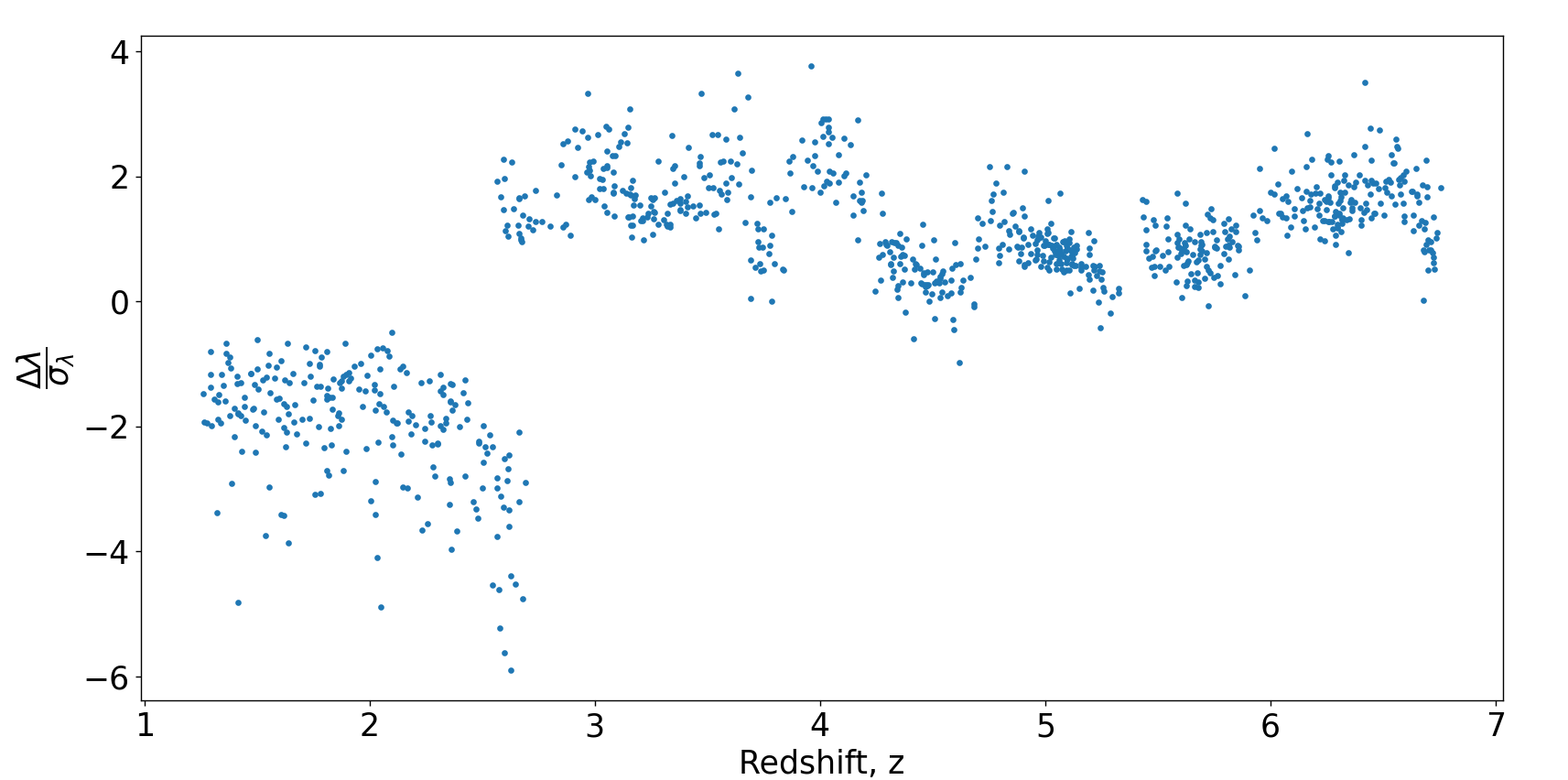}
    \caption{Redshift dependence of the scale of deviation of the KSP value from the generated distribution.}
    \label{fig:dlambda_simga}
\end{figure}

To estimate the KSP, the wavelength values in each interval were normalized to the mean value ($0)$ and the variance (1). Normalization enables a generalized normal distribution to be used as the theoretical distribution, with a single free parameter representing the sharpness of the distribution. The generalized normal distribution with variance equal to $1$ was taken as the theoretical distribution. Figure~\ref{fig:ksp_z} shows the dependence of the KSP value on the redshift in this scheme and its comparison with the KSP of the mock data of the same distribution. To illustrate the scale of deviation of the KSP of the real data from that of the generated distribution sample, the $\frac{\Delta\lambda}{\sigma_\lambda}$ dependence on $z$ is plotted in Fig.~\ref{fig:dlambda_simga}, where $\Delta \lambda$ is the difference between the the mean values of the KSP and $\sigma_\lambda$ is the standard deviation of the mock data KSP. The plot shows that the deviation reaches a level of 4$\sigma$.

\begin{figure}[h!]
    \centering
    \includegraphics[width=\linewidth]{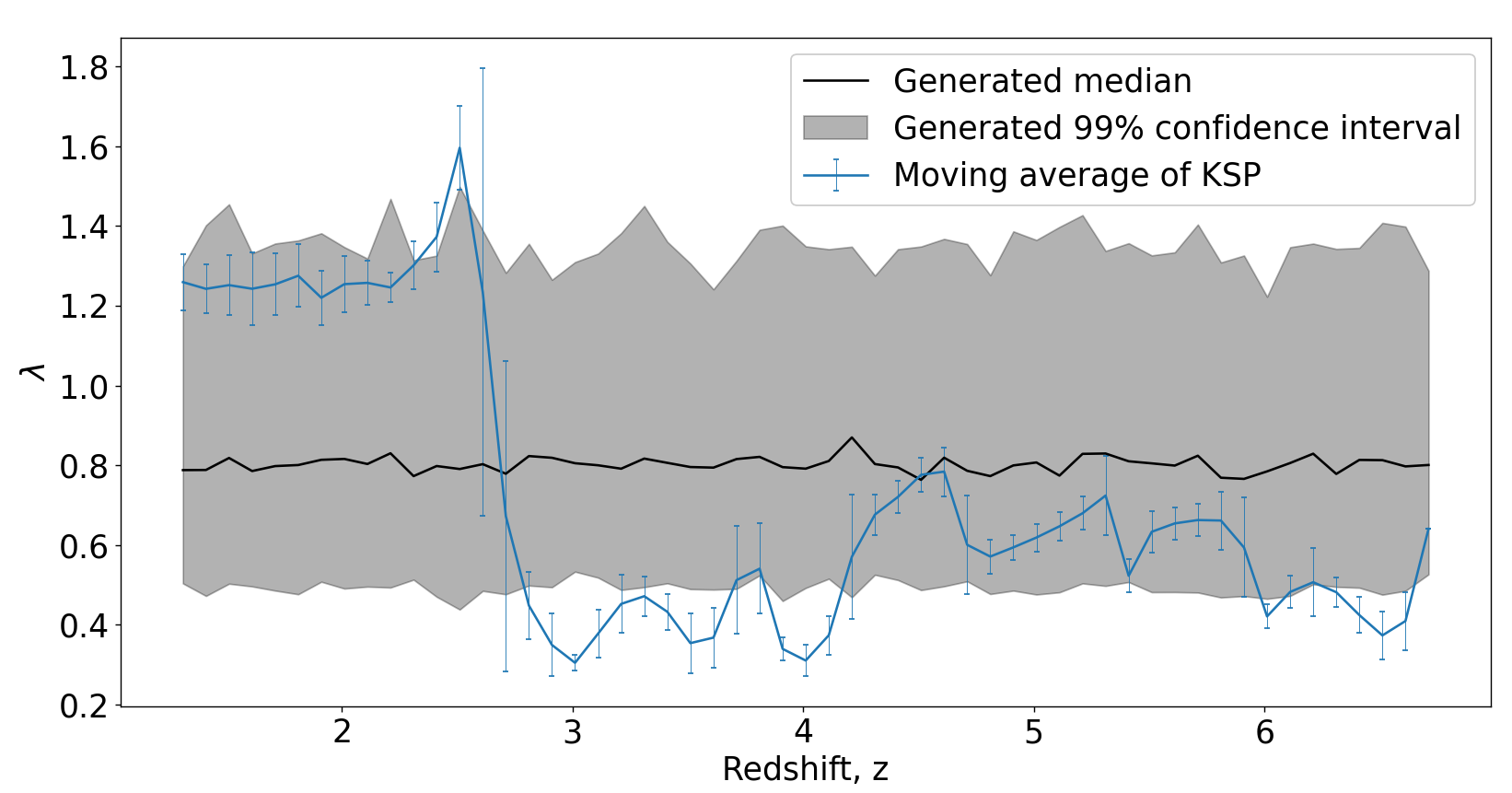}
    \caption{Moving average of Fig.~\ref{fig:ksp_z} with window $\Delta z =0.1$. Blue points are the averaged values, and the error bars are the standard deviations of the points inside the window. The black line denotes the averaged median KSP of the generated distribution, and the shaded region is the 99\% confidence interval around the median.}
    \label{fig:ksp_z_ma}
\end{figure}

The resulting dependence in Fig.~\ref{fig:ksp_z} was smoothed using a moving average with a redshift window of $\Delta z = 0.1$ to provide a more robust evaluation of the deviation level. Figure~\ref{fig:ksp_z_ma} shows the averaged KSP values (with a 99\% confidence interval) of the generated distribution. A change in the KSP values is evident at $z \simeq 2.7$.

\section{Conclusions}

We have analyzed JWST deep galaxy spectral survey data by means of the Kolmogorov technique. The Kolmogorov stochastic parameter has been shown \citep{N,UMN} to enable comparisons of the tiny randomness properties of cumulative signals, that is, signals composed of regular and random (stochastic) sub-signals. We analyzed the spectral data of the JWST galaxy survey up to redshift $z\simeq 7$ to KSP-test whether the data of the same survey keep certain properties independent of the redshift, or if they undergo distortions or evolve. Thus, it is crucial that the instrumental noise and certain systematics are identical for the galaxies of the dataset.

The results of the KSP analysis show a change in the JWST galaxy spectral properties at $z \simeq 2.7$, at over a 99\% confidence level. Namely, the results indicate a change in the random and regular components of a given galactic signal at a certain redshift, that is, the appearance or disappearance of certain agents contributing to that spectral signal at that redshift. The nature of this effect can be related to the features of the galaxy evolution and/or to the properties of the intergalactic medium through which the photon beams are propagating, linked with, for example, galactic dark haloes \citep{Paol} and their broad extensions \citep{Mis}, or an invisible population of galaxies \citep{Var}. Therefore, studying the revealed galactic spectral signature variation at certain redshifts by other means and using other data samples within broad redshift intervals can shed light on its nature.

\begin{acknowledgements}
We are thankful to the referee for valuable comments. We acknowledge the use of NASA/ESA/CSA \textit{James Webb} Space Telescope
data obtained from the Mikulski Archive for Space Telescopes at the Space Telescope Science Institute. 
\end{acknowledgements}






\label{lastpage}
\end{document}